\documentclass[floatfix, twocolumn, 10pt, aps, prd, showpacs, amsmath, amssymb, superscriptaddress]{revtex4-1}
\usepackage{graphicx}
\usepackage{hyperref}
\usepackage{subeqnarray}
\usepackage{color}
\usepackage[normalem]{ulem}
\usepackage{url}
\usepackage{multirow}
\usepackage{array}
\hypersetup{colorlinks=false}
\hypersetup{linktocpage}
\definecolor{darkblue}{RGB}{16,78,139}
\definecolor{darkgreen}{rgb}{0.04, 0.7, 0.2}
\newcommand{\beq}{\begin{equation}}
\newcommand{\eeq}{\end{equation}}
\newcommand{\beqn}{\begin{eqnarray}}
\newcommand{\eeqn}{\end{eqnarray}}
\newcommand{\beqs}{\begin{subeqnarray}}
\newcommand{\eeqs}{\end{subeqnarray}}
\newcommand{\nn}{\nonumber}

\begin{document}
\title{Ergoregion instability of a rotating quantum system}

\author{Leandro A. Oliveira}
\email{laoliveira@ufpa.br}
\affiliation{Campus Salin\'opolis, Universidade Federal do Par\'a, 68721-000, Salin\'opolis, Par\'a, Brazil}
\affiliation{Faculdade de F\'{\i}sica, Universidade Federal do Par\'a, 66075-110, Bel\'em, Par\'a, Brazil}

\author{Luis J. Garay}
\email{luisj.garay@ucm.es}
\affiliation{Departamento de F\'isica Te\'orica,
Universidad Complutense de Madrid, 28040 Madrid, Spain\\ and
Instituto de Estructura de la Materia (IEM-CSIC), Serrano 121, 28006 Madrid, Spain}

\author{Lu\'is C. B. Crispino}
\email{crispino@ufpa.br}
\affiliation{Faculdade de F\'{\i}sica, Universidade Federal do Par\'a, 66075-110, Bel\'em, Par\'a, Brazil}

\date{\today}

\begin{abstract}
Using the analogy between acoustic perturbations in an ideal fluid and the description of a Klein-Gordon scalar field in a curved spacetime, we study the quasinormal modes of a quantum system: the rotating Bose-Einstein condensate. To compute quasinormal frequencies, we use two different numerical techniques, namely the direct integration and the continued-fraction methods. We study in detail the ergoregion instability of this linearly perturbed system, comparing the results with different setup configurations.
\end{abstract}


\maketitle

\section{Introduction}
Acoustic analogue systems have shown to be interesting alternatives to investigate (theoretically and experimentally) some properties of compact objects~\cite{Unruh:1980cg,Visser:1997ux,AMproc,Barcelo:2005fc,Visser:2001fe}, as black holes, which so far cannot be subject to experimental reproduction in laboratory. Among the properties that can be studied in acoustic analogue systems, stand out: absorption and scattering of waves~\cite{Crispino:2007zz, Oliveira:2010zzb, Dolan:2009zza, Dolan:2011zza}, quasinormal modes~\cite{Berti:2004ju, Cardoso:2004fi, Dolan:2010zza, Dolan:2011ti}, ergoregion instability~\cite{Oliveira:2014oja, Oliveira:2015vqa, Oliveira:2016adj, OC:2018}, and stationary configurations~\cite{Hod:2014hda, Benone:2014nla}. Acoustic analogues have been studied in various systems, among which we can mention the perfect fluids and Bose-Einstein condensates (BECs)~\cite{Garay:1999sk, Garay:2000jj, Barcelo:2003wu}, the latter being shown as a fruitful proposal of an experimental setup. Recently an experimental apparatus in BECs was used for attempts to observe some aspects of Hawking radiation in acoustic analogue systems~\cite{Steinhauer:2014dra, Steinhauer:2015ava}.

In this work we investigate the instability of a rotating BEC. Previous studies on the instability of acoustic analogue systems were performed for incompressible~\cite{Oliveira:2014oja} and for compressible~\cite{Oliveira:2015vqa, Oliveira:2016adj, OC:2018} (perfect) fluids. As essential tools for the investigation of instabilities, there are the quasinormal modes (QNMs) of a system. QNMs are associated with purely outgoing modes at spatial infinity, each mode being described by a complex frequency~\cite{Berti:2009kk, Kokkotas:1999bd, Nollert:1999ji}. As a purely circulating system, we describe the effective spacetime of a compressible hydrodynamic vortex~\cite{Cherubini:2011zza, Oliveira:2015vqa}, compatible with an experimental setup in a rotating BEC. QNMs were previously studied in one-dimensional flows in a BEC with step-like discontinuity~\cite{Barcelo:2007ru}. Furthermore, the scattering of the sound waves was studied for a hydrodynamic vortex with a density profile of a BEC~\cite{Slatyer:2005ty}. 

The remainder of this paper is structured as follows. In Sec.~\ref{sec-spacetimes} we describe acoustic spacetimes in the context of a rotating compressible fluid. In Sec.~\ref{sec-BEC} we describe a purely circulating BEC as a rotating acoustic analogue. In Sec.~\ref{sec-Perturbations} we study the propagation of linear perturbations in this compressible system, using the description in the frequency domain. In Sec.~\ref{sec-Numerical} we obtain the QNM frequencies of this system using two different methods: direct integration (DI) and continued-fraction (CF). In Sec.~\ref{sec-Results} we investigate the ergoregion instability of the BEC hydrodynamic vortex, validating and commenting our results, comparing the QNM frequencies obtained via DI and CF methods. We conclude with a brief discussion in Sec.~\ref{sec-Conclusion}\,.

\section{Effective spacetimes of rotating acoustic analogues}
\label{sec-spacetimes}
Requiring a fluid flow to be irrotational (i.e., with zero vorticity), namely
\beq
\nabla \times \vec{v}=0,
\label{Irrot_1}
\eeq
where $\vec{v}$ is the flow velocity, we may write
\beq
\vec{v}=-\nabla \Phi,
\label{Irrot_2}
\eeq
where $\Phi$ is the velocity potential.

We may describe the irrotational flow of an inviscid fluid without thermal conductivity (ideal fluid) using the Euler equation, as follows
\beq
\frac{\partial \vec{v}}{\partial t} +\frac{1}{2}\nabla|\vec{v}|^2 +\frac{\nabla P}{\rho}+\nabla V_{\rm ext}=0,
\label{Euler_1}
\eeq
where $V_{ext}$ is an external potential, $P$ is the pressure and $\rho$ is the mass density of the fluid.  We consider that the ideal fluid is barotropic, i.e., with an equation of state (EoS) such that:
\beq
P=P\left(\rho \right).
\label{Baro}
\eeq

Furthermore, we impose mass conservation, which may be described by the continuity equation, namely
\beq
\frac{\partial \rho}{\partial t}+\nabla \cdot\left(\rho \vec{v}\right)=0.
\label{Cont_1}
\eeq

Using the description for effective curved spacetimes in ideal fluids proposed by Unruh~\cite{Unruh:1980cg}, we may obtain from Eqs.~\eqref{Irrot_2}--\eqref{Cont_1} that the propagation of linear disturbances in the background flow can be governed by the Klein-Gordon equation~\cite{Visser:1997ux}, namely
\beq
 \nabla_\mu \nabla^\mu \phi = \frac{1}{\sqrt{|g|}} \partial_\mu \left( \sqrt{|g|} g^{\mu \nu} \partial_\nu \phi \right) = 0 , \label{Klein}
\eeq
with $g^{\mu \nu}$ being the contravariant \emph{effective metric}, $g \equiv \det(g_{\mu \nu})$, and 
$\phi$ is the velocity potential associated to the linear  perturbations, i.e.
\beq
\delta\vec{v}= -\nabla \phi, 
\label{v-pert}
\eeq
where $c_{\rm s}$ is the speed of sound, which can be written as
\beq
c_{\rm s}\equiv\sqrt{dP/d\rho}.
\label{Speed_sound_1}
\eeq

An appropriate way to describe the properties of the effective spacetime of a rotating acoustic analogue is adopting a cylindrical coordinate system $(t, r, \theta, z)$. 
Here we focus on the hydrodynamic vortex as a rotating acoustic analogue -- a purely circulating fluid -- whose flow velocity $\vec{v} = v_r \hat{r}+v_\theta \hat{\theta}+v_z \hat{z}$ is such that both the radial $v_{r}$ and the z-component $v_{z}$ vanish, so that
\beq
\vec{v}=v_\theta \hat{\theta}.
\label{Flow_velocity_1}
\eeq
Furthermore, from Eq.~\eqref{Irrot_1}, we find that the velocity flow may be written as
\beq
\vec{v}=\frac{C}{r} \hat{\theta},
\label{Flow_velocity_2}
\eeq
where $C$ is a constant related with the conserved circulation.

From the Klein-Gordon equation, given by Eq.~\eqref{Klein}, we may write, in cylindrical coordinates, the contravariant effective metric $g^{\mu \nu}$ for a hydrodynamic vortex, namely
\beqn
 g^{\mu \nu} =-\frac{1}{\rho c_{\rm s}}\left[
\begin{array}{cccc}
1 & 0 & v_\theta/r & 0\\
\nn\\
0 & -c_{\rm s}^2 & 0 & 0\\
\nn\\
 v_\theta/r & 0 & (v_\theta^2-c_{\rm s}^2)/r^2 & 0\\
\nn\\
0 & 0 & 0  &-c_{\rm s}^2
\end{array}
\right].\nn\\
\label{Contra_metric}
\eeqn

Thus, the line element $ds^2=g_{\mu \nu}dx^\mu dx^\nu$ of a hydrodynamic vortex can be written as~\cite{Visser:2004zs,Fischer:2001jz,Marecki, Marecki:2011ve}
\beqn
ds^2=\frac{\rho}{c_{\rm s}}\left[-c_{\rm s}^2dt^2+\left(rd\theta -\frac{C}{r} dt \right)^2+dr^2+dz^2\right].\nn\\
\label{Elem_vortex}
\eeqn
Note that the line element~\eqref{Elem_vortex} may be written as a function of a quantity only, namely the fluid density $\rho$, since the speed of sound explicitly depends on the density [as may be seen from Eq.~\eqref{Speed_sound_1}]. Thus, to study the properties of the hydrodynamic vortex it is necessary to know the fluid density profile. Here we consider that the density $\rho$ and the speed of sound $c_{\rm s}$ are functions of the radial coordinate only, obtained from local properties of the unperturbed fluid flow~\cite{Visser:1997ux}. Furthermore, as can be seen from the line element~\eqref{Elem_vortex}, the spacetime of the hydrodynamic vortex has no event horizon, but has an ergoregion delimited by an outer boundary $r_{\rm e}$, which can be obtained from~\cite{Visser:1997ux, Oliveira:2014oja, Oliveira:2015vqa}
\beq
c_{\rm s}(r=r_{\rm e})^2 = |\vec{v}(r=r_{\rm e})|^2.
\label{re_1}
\eeq

Next, we will obtain the expressions for the density $\rho$ and the speed of sound $c_{\rm s}$, as functions of $r$, for the hydrodynamic vortex with a compatible experimental setup in a BEC.

\section{BEC as a rotating acoustic analogue}
\label{sec-BEC}
The BEC considered in this work is described as a physical system obtained when a set of bosonic particles, which are subjected to an exterior potential and interact in pairs with neighbor particles, occupy the same quantum state - the ground quantum state of the system~\cite{Pethick, Pitaesvskii, Garay:1999sk, Garay:2000jj, Barcelo:2003wu}. This system is formed at the limit in which the temperature of the bosons is decreased to extremely low values ($T\approx0$). At zero temperature, all particles of the condensate occupy the quantum ground state. This system can be described by a wave function, whose evolution is governed by the Gross-Pitaevskii equation~\cite{Pethick, Pitaesvskii}, namely
\beq
i\hbar \frac{\partial \Psi}{\partial t}=\left(-\frac{\hbar^2}{2M} \nabla^2 +V_{\rm ext}\left(\vec{r}\right) + U |\Psi|^2\right)\Psi,
\label{gross}
\eeq
where $\hbar$ is the reduced Planck constant, $V_{\rm ext}\left(\vec{r}\right)$ is an external potential, $\Psi$ is a quantum field that describes the BEC, $M$ is the individual mass of each boson and $U$ parameterizes the strength of the interaction between bosons~\cite{Pethick, Pitaesvskii}, as
\beq
U=\frac{4\pi a \hbar^2}{M},
\eeq
with $a$ being the scattering length between two bosons in the condensate~\cite{Pethick, Pitaesvskii}. 

The total number of bosons in the condensate is given by $N = \int dx^3 |\Psi|^2$ and the density of bosons of the BEC can be written as
\beq
\rho=|\Psi|^2.
\eeq

We may rewrite the Gross-Pitaevskii equation using the Madelung representation for the wave function $\Psi$, namely
\beq
\Psi=\sqrt{\rho}\exp\left[ {-\frac{i}{\hbar}}\left(\Phi\left(t,\vec{r}\right)+\mu_{\rm c} t \right) \right],
\label{madelung}
\eeq
where $\rho$ is the density of the Madelung fluid~\cite{Barcelo:2003wu} and $\mu_{\rm c}$ is the chemical potential (here we consider $\mu_{\rm c}$ fixed~\cite{Pethick, Pitaesvskii}). 

Substituting Eq.~(\ref{madelung}) in the Gross-Pitaesvskii equation~(\ref{gross}), we obtain the following expression
\beqn
&-&\frac{i\hbar \rho^{-\frac{1}{2}}}{2} \frac{\partial \rho}{\partial t}-\sqrt{\rho} \frac{\partial \Phi}{\partial t}-\frac{\hbar^2}{2M}\nabla^2 \sqrt{\rho}+\frac{i\hbar}{M}\left(\nabla \sqrt{\rho}\right) \cdot \left(\nabla \Phi \right)\nonumber\\
\nonumber\\
&+&\frac{\sqrt{\rho}}{2M} \left(\nabla \Phi \right) \cdot  \left(\nabla \Phi \right)+
\frac{i\hbar \sqrt{\rho}}{2M}\nabla^2 \Phi+ V_{\rm ext} \sqrt{\rho}+U \rho \sqrt{\rho}\nonumber\\
\nonumber\\
&-&\mu_{\rm c} \sqrt{\rho}=0. 
\label{re_im_gross}
\eeqn
Separating the real and imaginary parts of Eq.~(\ref{re_im_gross}), we find the following expression for the real part:
\beqn
&-&\frac{\partial}{\partial t}\left(\frac{\Phi}{M}\right)+\frac{1}{2}\nabla \left(-\frac{\Phi}{M}\right) \cdot \nabla \left(-\frac{\Phi}{M}\right)\nn\\
&+&\frac{U \rho}{M}+\frac{V_{\rm ext}}{M}+\frac{V_{\rm Q}}{M}-\frac{\mu_{\rm c}}{M}=0
\label{boseul}
\eeqn
and for imaginary part
\beq
\frac{\partial \rho}{\partial t}+\nabla \cdot\left[\rho \nabla \left(-\frac{\Phi}{M}\right)\right]=0,
\label{boscon}
\eeq
where 
\beq
V_{\rm Q} \equiv -\frac{\hbar^2}{2M \sqrt{\rho}}\nabla^2 \sqrt{\rho}
\label{quantum_pot}
\eeq
is the so-called quantum potential. This quantum potential can be neglected when the Thomas-Fermi approximation is considered~\cite{Pitaesvskii}. Note that, as it can be seen from Eq.~\eqref{quantum_pot}, if the density is sufficiently small (and the term $\nabla^2 \sqrt{\rho}$ is large enough), the quantum potential cannot be neglected~\cite{Pitaesvskii}.  

Essentially, Eqs.~\eqref{boseul} [more precisely, the gradient of Eq.~\eqref{boseul}] and~(\ref{boscon}), under Thomas-Fermi approximation, can be rewriten, respectively, as the Euler equation [Eq.~\eqref{Euler_1}] and the continuity equation [Eq.~\eqref{Cont_1}], where the flow velocity is $\vec{v}=-\nabla \left(\Phi/M\right)$, what implies that the BEC is irrotational. 

Furthermore, comparing Eqs.~\eqref{Euler_1} and~\eqref{boseul}, we obtain a relation between pressure and density, defining then an EoS for the BEC, namely
\beq
P=\frac{U \rho^2}{2M},
\label{BEC_EoS}
\eeq
which denotes that the BEC (in the Thomas-Fermi approximation) can be represented as an ideal fluid, which is barotropic and irrotational.

From Eq.~\eqref{Speed_sound_1} and the EoS~\eqref{BEC_EoS}, we may write the speed sound $c_{\rm s}$ in the BEC as
\beq
c_{\rm s} = \sqrt{\frac{4\pi a \hbar^2 \rho}{M^2}}.
\label{BEC_Speed_1}
\eeq

To describe a BEC hydrodynamic vortex, we recall that the flow velocity has only an angular component, given by
\beq
v_\theta=\frac{C}{r},
\label{BEC_Velo}
\eeq
with 
\beq
C \equiv \dfrac{\ell\hbar}{M}, 
\eeq
where $\ell$ is an integer number associated to the quantization of the circulation of the BEC hydrodynamic vortex~\cite{Pethick, Pitaesvskii}.

The density profile of the BEC may be determined directly from Eq.~\eqref{boseul} [in the Thomas-Fermi approximation], being
\beqn
\rho =\frac{M}{4 \pi a \hbar^2}\left[\mu_{\rm c}-V_{\rm ext}-\frac{\hbar^2 \ell^2}{2 M r^2}\right].
\label{boseul1}
\eeqn

Before proceeding to the derivation of a physically acceptable expression for the density $\rho$ of the BEC hydrodynamic vortex, some considerations about the external potential $V_{\rm ext}$ figuring in Eq.~\eqref{boseul1} are in order. In this paper, we consider a constant external potential, namely $V_{\rm ext} = V_{0}=\text{\it constant}$~\cite{Pethick, Pitaesvskii}. 

From Eq.~\eqref{boseul1}, we may write an expression for the density profile of a vortex in a uniform medium, namely
\beq
\rho=\rho_\infty \left(1-\frac{r_{\rm c}^2}{r^2} \right),
\label{BEC_Density_1}
\eeq
with 
\beq
\rho_\infty\equiv \frac{M \left(\mu_{\rm c} -V_{0}\right)}{4 \pi a \hbar^2},
\label{Density_inf}
\eeq
where $\rho_\infty$ is density of the BEC at $r \rightarrow \infty$ (at large distances from the center of the vortex), and $r_{\rm c}$ is the so-called critical radius, given by
\beq
r_{\rm c} \equiv \frac{|\ell|}{\sqrt{8 \pi a \rho_\infty}},
\label{rc_1}
\eeq
defining the position where the density of the vortex goes to zero. The quantities $\rho_\infty$ and $a$ can be determined from the experimental setup~\cite{Slatyer:2005ty}. Furthermore, from Eq.~\eqref{Density_inf}, the chemical potential $\mu_{\rm c}$ can be obtained, namely
\beq
\mu_{\rm c}=\frac{4\pi a \hbar^2 \rho_\infty}{M}+V_{0}.
\eeq

Note that the critical radius $r_{\rm c}$ may be interpreted as a characteristic length that delimits where the Thomas-Fermi approximation is valid~\cite{Pitaesvskii}, i.e., for sufficiently large distances from $r = r_{\rm c}$ [recalling that the density reaches its minimum value ($\rho \rightarrow 0$) at $r \rightarrow r_{\rm c}$], the quantum potential $V_{\rm Q}$ is small enough and can be neglected [cf. Eq.~\eqref{quantum_pot}]. 

It is worth noting that the corresponding Kretschmann invariant~\cite{Wald} goes to infinity at the critical radius $ r_{\rm c}$, denoting that this point is an essential singularity~\cite{Oliveira:2015vqa,Cherubini:2011zza}.

From Eqs.~(\ref{BEC_Speed_1}) and~\eqref{BEC_Density_1}, we may write the speed of sound as 
\beq
c_{\rm s} = c_{s \infty}\sqrt{1-\frac{r_{\rm c}^2}{r^2}},
\label{BEC_Speed_2}
\eeq
where $c_{s\infty} = \sqrt{4\pi a \hbar^2 \rho_{\infty}/M^2}$ is the speed of sound at infinity. The speed of sound, as well as the density profile [cf. Eq.~\eqref{BEC_Density_1}], vanishes at the critical radius $r_{\rm c}$.

Using Eqs.~\eqref{re_1},~\eqref{BEC_Velo} and~\eqref{BEC_Speed_2}, we may obtain the outer boundary of the ergoregion, namely
\beq
r_{\rm e}=\frac{\sqrt{3}|\ell|  }{\sqrt{8 \pi a \rho_\infty }}.
\label{BEC_re_2}
\eeq
Comparing Eqs.~\eqref{rc_1} and~\eqref{BEC_re_2}, we find a relation between $r_{\rm e}$ and $r_{\rm c}$, as follows
\beq
\frac{r_{\rm e}}{r_{\rm c}}=\sqrt{3},
\eeq
denoting that the ergoregion encompasses the location where the density of the vortex vanishes, i.e., $r_{\rm e}>r_{\rm c}$.

\subsection{On the validity of the Thomas-Fermi approximation}
\label{sec-validity}

We have seen that under certain approximations, namely in the regime in which we can ignore the quantum potential, the background density is given by Eqs. \eqref{BEC_Density_1}--\eqref{rc_1} and furthermore, that the (phase) perturbations obey an effective Klein-Gordon equation in a curved spacetime with metric \eqref{Contra_metric}. In this section we will briely analyze the conditions under which this effective formulation provides an accurate description of sound propagation in the vortex BEC.

As we have already discussed, the effective Klein-Gordon equation can be obtained by perturbing Eqs. \eqref{boseul} and \eqref{boscon}, i.e. by expanding these equations to linear order in the phase pertubation $\phi=\delta \Phi$ and the density perturbaton $\varrho=\delta\rho$, solving the first equation for $\varrho$ in terms of $\phi$, and replacing the result in the second. In order to do so, we must ensure that the contribution from the quantum potential is negligible as compared with the interaction term. This amounts to the condition
\beq
U\varrho\gg \frac{\hbar^2}{4M\sqrt{\rho }} \bigg|\nabla^2\bigg(\frac{\varrho}{\sqrt{\rho }}\bigg)-\frac{\varrho}{\rho}\nabla^2\sqrt{\rho }\,\bigg|.
\eeq
From the second term we obtain the Thomas-Fermi approximation for the background density. This implies [using the lowest order expression \eqref{BEC_Density_1}] that if we want it to be $\epsilon$ times smaller than the interaction term, we need that
\beq 
 \frac{r-r_{\rm c}}{r_{\rm c}}> \frac1{(16 \ell^{2}\epsilon)^{1/3}},
\eeq
The first term gives the following upper bound for the wave number $k$ of the perturbations:
\beq
k\ll \ell/r_{\rm c}=\sqrt{8\pi a\rho_\infty}.
\eeq

Therefore, the effective Klein-Gordon description is only valid well beyond the central core of the vortex for low $\ell$, while for large $\ell$, the Thomas-Fermi approximation is valid at much closer distances. Even for low $\ell$, the analysis of the quasinormal modes for this type of vortices within the metric approximation is still interesting, among other things, because of the role played by the boundary conditions that have to be imposed close to the center.

\section{Perturbations in a BEC hydrodynamic vortex}
\label{sec-Perturbations}
Since the spacetime of the hydrodynamic vortex has a cylindrical symmetry, we may denote the angular dependence of the field $\phi$ as $e^{i m \theta}$ [cf. Eq.~\eqref{v-pert}], where $m$ is an integer number called azimuthal number, which is related to the angular momentum of the perturbation. Furthermore, we may describe the propagation of the perturbation in the frequency domain, assuming that the field $\phi$ depends on time as $e^{i \omega t}$, with $\omega$ being the frequency of the perturbation. Then, the field $\phi$ can be written as follows
\beq
\phi(t, r, \theta, z) = \frac{1}{\sqrt{r}} \sum_{m=-\infty}^{\infty} u_{\omega m}(r) \exp{\left[ i\left( m \theta -\omega t \right)\right] },
\label{modesum}
\eeq
where $u_{\omega m}(r)$ represent the single-frequency modes, i.e., the perturbations described in the frequency domain.

Using the covariant metric components $g^{\mu \nu}$, given by Eq.~\eqref{Contra_metric}, and Eq.~\eqref{modesum} together with Eq.~\eqref{Klein}, we obtain the following ordinary differential equation: 
\beqn
&&\left[\rho\frac{d}{dr}\left(\rho\frac{d}{dr}\right)+\frac{\rho^2}{c_{\rm s}^2}\left(\omega-\frac{Cm}{r^2} \right)^2 \right. \nn\\
&&\left.-\frac{\rho^2}{r^2}\left(m^2-\frac{1}{4} \right)-\frac{\rho}{2r}\frac{d\rho}{dr}\right]u_{\omega m}(r)=0.
\label{radial}
\eeqn

We may substitute Eqs.~\eqref{BEC_Density_1} and~\eqref{BEC_Speed_2} into Eq.~\eqref{radial}, to obtain the following ordinary differential equation for a BEC hydrodynamic vortex
\beqn
&&\left[ x^2\left(3x^2-1 \right)\frac{d^2 }{dx^2}+2x\frac{d }{dx}+\left(\sqrt{3} \varpi x^2 -\frac{\sqrt{2}mC}{|C|} \right)^2\right.\nn\\
&&\left.-\left(3x^2-1 \right)\left(m^2-\frac{1}{4} \right)-1\right]u_{\varpi m}(x)=0,
\label{Edo_BEC}
\eeqn
where we defined a dimensionless frequency 
\beq
\varpi \equiv  \frac{\omega r_{\rm e}}{c_{{\rm s}\infty}},
\label{freq_dim}
\eeq
and a dimensionless radial coordinate
\beq
x \equiv \frac{r}{r_{\rm e}}.
\label{radial_dim}
\eeq
Equation~\eqref{Edo_BEC} has regular singular points at the origin and at the critical radius $x=x_{\rm c}$ ($x_{\rm c}\equiv r_{\rm c}/r_{\rm e}=1/\sqrt{3}$), and an irregular singular point at spatial infinity.

\section{Numerical methods}
\label{sec-Numerical}
In order to find the solutions of Eq.~\eqref{Edo_BEC}, we impose boundary conditions at a certain position $x=x_{\rm min}$ ($x_{\rm min}>x_{\rm c}$) and at spatial infinity $x\rightarrow \infty$. We may assume a boundary condition of Neumann type at $x=x_{\rm min}$ (close to the center of the vortex) such that
\beqn
\left[\frac{d}{dx}\left( \frac{u_{\varpi m}(x)}{\sqrt{x}}\right) \right]_{x=x_{\rm min}}=0\,.
\label{BCII}
\eeqn
The boundary condition~\eqref{BCII} can be physically interpreted as a cutt-off, close to the center of the vortex, on the radial velocity increment ($\delta\vec{v}= -\nabla \phi$) associated to the linear perturbation [cf. Eq.~\eqref{modesum}]. This is related to a rigid barrier which has a great opposition to this increment of the velocity of the fluid (large acoustic impedance) placed at $r = r_{\rm min}$.

For large radial distances, in accordance with the asymptotic behavior of Eq.~\eqref{Edo_BEC}, we may write the following solution
\beqn
u_{\varpi m}\left(x\to \infty \right) \sim \exp{\left( {i\varpi x}\right)}\,.
\label{Boun_2}
\eeqn

We employed two numerical methods in the frequency domain to solve directly the ordinary differential equations~\eqref{Edo_BEC}, using the boundary conditions previously proposed, in order to obtain the frequency spectra $\varpi$ for different values of the azimuthal number $m$ and of $x_{\rm min}$.

\subsection{Direct integration method} 
\label{sec-Direct}
We may obtain the QNM frequencies from Eq.~\eqref{Edo_BEC} determining their solutions via the direct integration (DI) method~\cite{Dolan:2010zza}. This procedure may be implemented as follows:
\begin{itemize}
\item[(i)] We impose boundary conditions on the wave function $u_{\varpi m}\left(x \right)$ and its derivative at spatial infinity, namely
\beqn
u_{\varpi m}\left(x\to \infty \right) &=& \exp\left( {i\varpi x}\right) \sum_{j=0}{\frac{b_j}{x^j}}\,,
\label{serie2}
\label{dserie2}
\eeqn
where $b_j$ are coefficients which can be determined by collecting the same inverse powers of $x$ from Eq.~\eqref{Edo_BEC} at $x\rightarrow \infty$, keeping fixed the parameters $m$ and $\omega$. For example, 
\beq
b_1=\frac{i \left(-3+12 m^2+8 \sqrt{6} m \omega -4 \omega ^2\right)}{24 \omega }.\nn
\eeq

\item[(ii)] We integrate inwards Eq.~\eqref{Edo_BEC}, in the range $\infty > x \geq x_{\rm min}$.

\item[(iii)] At $x=x_{\rm min}$, we extract the QNM frequencies as roots of $\left[\frac{d}{dx}\left( \frac{u_{\varpi m}(x)}{\sqrt{x}}\right) \right]_{x=x_{\rm min}}=0$, using a standard root-finding algorithm such as Newton's method.
\end{itemize}
%
\subsection{Continued fraction method} 
\label{sec-Continued}
Another way to obtain the QNM frequencies from Eqs.~\eqref{Edo_BEC} consists in representing the wave function $u_{\varpi m}(x)$ as a Frobenius-like power series around $x=x_{\rm min}$ that satisfies the boundary conditions~\eqref{BCII} and~\eqref{Boun_2}, as follows
\beq
u_{\varpi m}(x) = \exp\left(i\varpi x\right) \sum_{n=0} a_n \left( 1 -\frac{x_{\rm min}}{x} \right)^{n}\,.   
\label{Ans_1}
\eeq
Substituting Eq.~\eqref{Ans_1} into Eq.~\eqref{Edo_BEC}, we find the following five-term recurrence relation:
\beqs \label{Reco_1}
&&\alpha_0a_2+\beta_0a_1 +\gamma_0a_0= 0, \\
&&\alpha_1a_3+\beta_1a_2+\gamma_1a_1 +\delta_1a_0 = 0,  \\
&&\alpha_n a_{n+2}+\beta_na_{n+1}+\gamma_n a_{n}+\delta_{n}a_{n-1}+\epsilon_{n}a_{n-2} = 0, \nn\\ &&\mbox{for} \hspace{0.1cm} n \geq 2,
\eeqs
where the recurrence coefficients $\alpha_n$, $\beta_n$, $\gamma_n$, $\delta_{n}$ and $\epsilon_{n}$ are complex functions of the azimuthal number $m$ and $x_{\rm min}$, given by
\beqn
&&\alpha_n=-4 n \left(1+n\right) \left(-1+3 x_{\rm min}^2 \right),\nn
\\ \nn
&&\beta_n = 8 n \left[n \left(-2+3 x_{\rm min}^2\right)+i x_{\rm min} \left(1-3 x_{\rm min}^2\right) \omega \right], \nn
\\ \nn
&&\gamma_n = 5-3 x_{\rm min}^2-12 \left(-1+n\right) n \left(-2+x_{\rm min}^2\right)\\ \nn 
&&+12 m^2 \left(-1+x_{\rm min}^2\right)-8 i \left[1+2 \left(-1+n\right)\right] x_{\rm min} \omega \\ \nn
&&+8 \sqrt{6} m x_{\rm min}^2 \omega -4 x_{\rm min}^2 \omega ^2,\nn
\\ \nn
&&\delta_{n}=6+24 m^2-16 \left(-1+n\right)^2+8 i \left(-1+n\right) x_{\rm min} \omega,\\ \nn
&&\epsilon_{n}= -3-12 m^2-4 \left(-1+n\right)+4 \left(-1+n\right)^2,\nn
\eeqn
being obtained from Eqs.~\eqref{Edo_BEC}, together with Eqs.~\eqref{Ans_1} and~\eqref{Reco_1}\,.

Using a double Gaussian elimination (cf. Refs.~\cite{Onozawa:1995vu,Oliveira:2014oja}), from the five-term recurrence relation~\eqref{Reco_1} we may write the following three-term recurrence relation:
\beqn
\alpha_n a_{n+2}+\beta_na_{n+1}+\gamma_n a_{n} = 0, \hspace{0.5cm} \mbox{for} \hspace{0.1cm} n \geq 0. 
\label{Reco_2}
\eeqn
Using Eqs.~\eqref{Edo_BEC},~\eqref{Ans_1} and~\eqref{Reco_2}, it is possible to express analytically the recurrence coefficients $\alpha_n$, $\beta_n$, and $\gamma_n$, as functions of the parameters $m$, $\omega$ and $x_{\rm min}$.

Considering the boundary condition~\eqref{BCII}, Eqs.~\eqref{Ans_1} and~\eqref{Reco_2}, we may obtain the following continued-fraction (CF)~\cite{Leaver:1985ax}:
\beqn
1 -2i\varpi x_{\rm min} +\dfrac{2\gamma_1}{\beta_1-\dfrac{\alpha_1\gamma_2}{\beta_2-\dfrac{\alpha_2\gamma_3}{\beta_3-...}}}=0.
\label{Cont_frac_BCII}
\eeqn
Again, we may obtain the QNM frequencies from Eq.~\eqref{Cont_frac_BCII} for different values of the azimuthal number $m$ and $x_{\rm min}$, using a standard root-finding algorithm such as Newton's method.

\section{Ergoregion instability}
\label{sec-Results}
In this section we investigate the ergoregion instability of the BEC hydrodynamic vortex.

\begin{figure*}[htpb!]
   \includegraphics[width=0.5\textwidth]{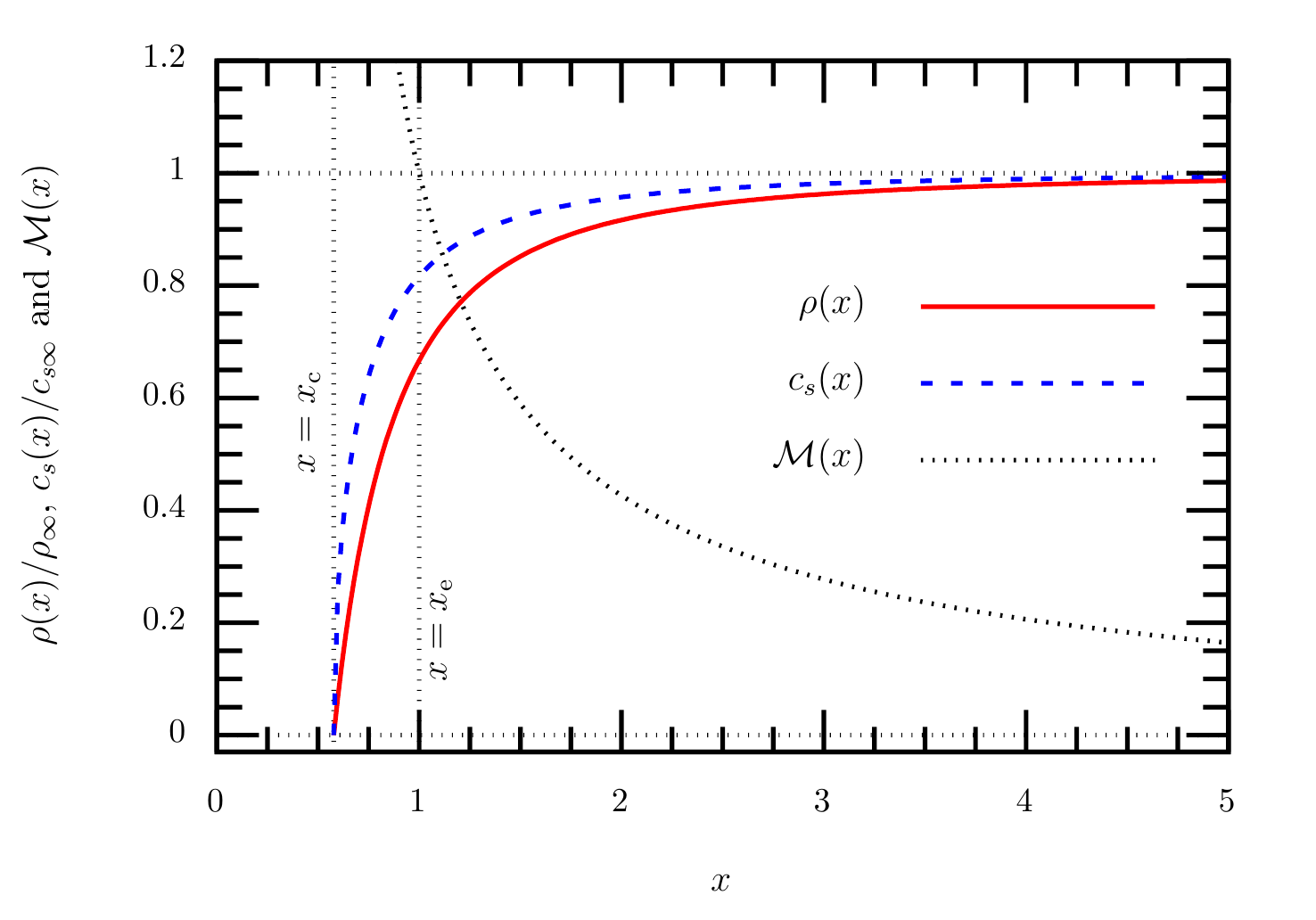}\includegraphics[width=0.5\textwidth]{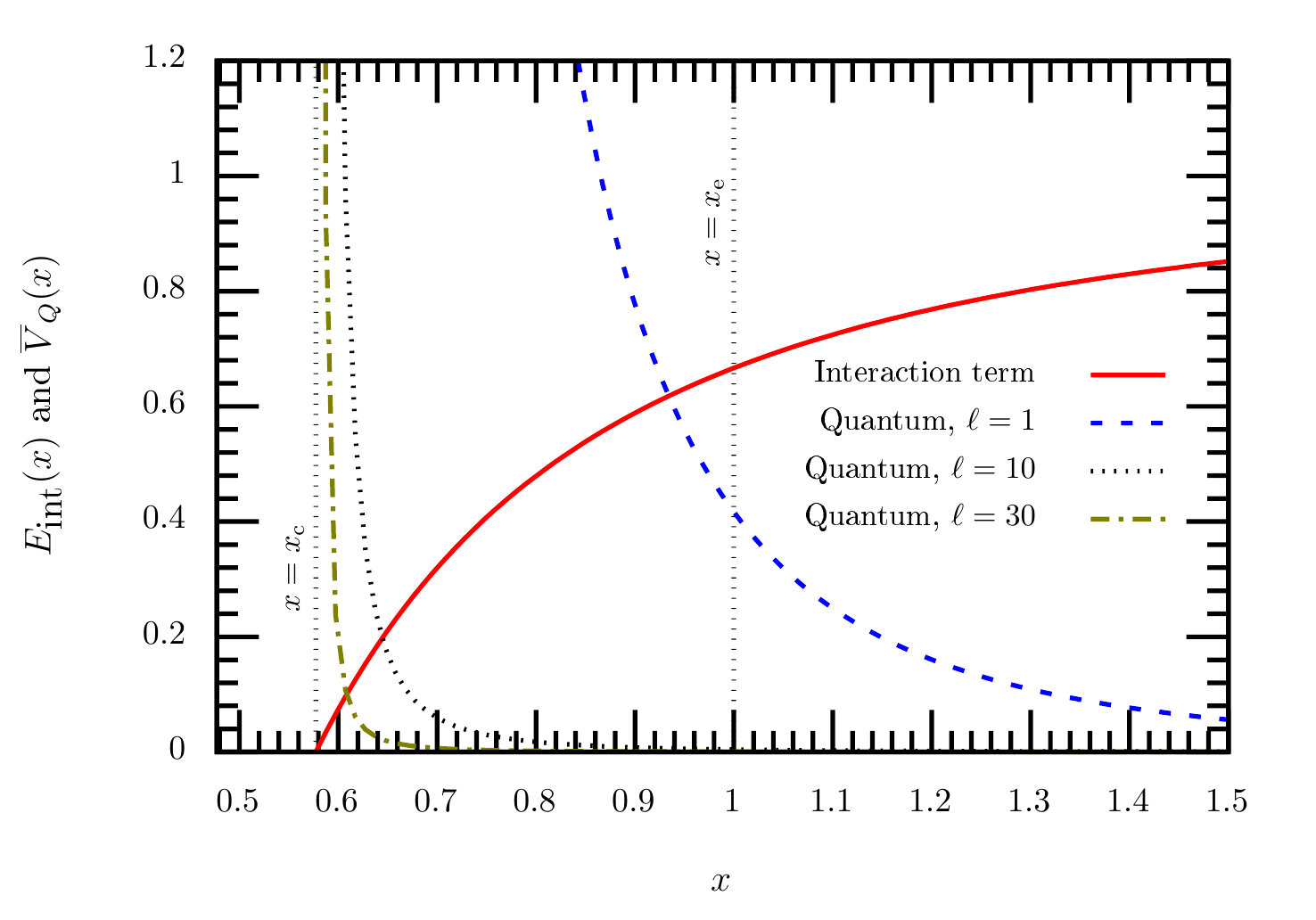}
   \caption{{\it Left:} Fluid density $\rho$, speed of sound $c_{\rm s}$, and local Mach number {\usefont{OMS}{cmsy}{m}{n} M}, for the BEC hydrodynamic vortex, as functions of $x$. We also exhibit the position of the critical radius, $x_{\rm c} \equiv 1/\sqrt{3}$, and of the outer boundary of the ergoregion, $x_{\rm e} \equiv 1$ (vertical dashed lines). {\it Right:} Dimensionless interaction term, $E_{\text{int}}(x)$ given by Eq.~(\ref{Eint}), and  dimensionless quantum potential, $\overline{V}_Q(x)$ given by Eq.~(\ref{VQbar}), for $\ell=1,\,10$ and $30$.}
   \label{fig-Density_Speed_BEC}
  \end{figure*}

In the left frame of Fig.~\ref{fig-Density_Speed_BEC} we plot the fluid density $\rho$ [given by Eq.~\eqref{BEC_Density_1}], the speed of sound $c_{\rm s}$ [given by Eq.~\eqref{BEC_Speed_2}] and the local Mach number {\usefont{OMS}{cmsy}{m}{n} M} $\equiv |\vec{v}|/c_{\rm s}$, as functions of $x$. In the right frame of Fig.~\ref{fig-Density_Speed_BEC} we plot the dimensionless interaction term
given by 
\beqn
E_{\text{int}} (x) \equiv \frac{U\rho(x)}{K_{\text{c}}}\, ,
\label{Eint}
\eeqn
and the dimensionless quantum potential 
\beqn
\overline{V}_Q(x) \equiv \frac{V_Q(x)}{K_{\text{c}}}\, ,
\label{VQbar}
\eeqn
with $K_{\text{c}} \equiv  4\pi a \hbar^2 \rho_\infty/M$. We note that as the density of the BEC decreases to zero at $x \rightarrow x_{\rm c}$, the quantum potential can assume a significant value [cf. Eq.~\eqref{quantum_pot}], being non negligible, and the Thomas-Fermi approximation is no longer valid in this case. Furthermore, we note that, the larger the value of $\ell$ is, the closer to the vortex the dimensionless quantum potential goes to zero. In our numerical simulations, we consider boundary conditions imposed at sufficiently large distances for the critical radius, e.g., for a range of values of $x_{\rm min}$ with $x_{\rm min}\geq 0.7$ (with $x_{\rm min}/x_{\rm c} > 1.2$).

Using two frequency domain methods (DI and CF methods), we computed the QNM frequencies. In order to verify the stability of the modes, boundary conditions of Neumann type are applied, outside and inside of the ergoregion, for this compressible acoustic system. 

For the QNM analysis, we assume the standard convention of ordering the imaginary part of the QNM frequencies $\varpi$~\cite{Berti:2009kk}. The fundamental mode ($n=0$) is the one with the largest imaginary part of the QNM frequencies. Thus, if the mode is unstable [${\rm Im}(\varpi)>0$], the fundamental mode corresponds to the smallest instability timescale, and for stable modes [${\rm Im}(\varpi)<0$], it corresponds to the longest-lived mode. 

From Eq.~\eqref{Edo_BEC}, it can be clearly seen that there are symmetries associated to the frequency $\varpi$, relating the co-rotating modes ($Cm>0$) and the counter-rotating ones ($Cm<0$), as follows
\beqn
&&\varpi({Cm>0}) = -\varpi^*({Cm<0}),
\label{Symm}
\eeqn
where ``$\,{}^*\,$'' denotes complex conjugation.  
From Eq.~\eqref{Symm}, we note that to each QNM frequency of a co-rotating mode there is a corresponding one of a counter-rotating mode with opposite real part and the same imaginary part~\cite{Note_2}. Henceforth, taking into account the symmetries~\eqref{Symm}, we may assume, without loss of generality, that $m > 0$ and $C > 0$. 

Estimates of the QNM frequencies $\varpi$ are exhibited in Table~\ref{tab-freq_1}, considering different values of the azimuthal number $m$ and applying the boundary condition~\eqref{BCII} at two different positions of the radial coordinate. These QNM frequencies are obtained via DI and CF methods~\cite{Note_1}. Note that, as the azimuthal number $m$ is increased, the magnitude of the real and imaginary parts of the QNM frequencies for the stable modes also increases, while the opposite happens for the unstable modes. 
The ergoregion instability can be more easily perceived for large values of $m$.
\begin{table}[htpb!]
\caption{QNM frequencies $\varpi$ of the BEC hydrodynamic vortex for different values of the azimuthal number $m$, for $x_{\rm min}=2.0$ (outside the ergoregion) and $x_{\rm min}=0.7$ (inside the ergoregion), obtained numerically from estimates via DI and CF methods~\cite{Note_1}. At $x=x_{\rm min}$, we imposed the boundary condition~\eqref{BCII}, and, at spatial infinity, we considered the boundary condition~\eqref{Boun_2}.}
\begin{center}
\begin{tabular*}{0.4\textwidth}{ c c @{\extracolsep{\fill} } c c}
\hline\hline
  \multicolumn{4}{c}{$x_{\rm min}=2.0 $ (outside the ergoregion)}  \\
\hline
 \multicolumn{1}{c}{$m$} & \multicolumn{1}{c}{Method}& \multicolumn{1}{c}{$\hbox{Re}(\varpi) $} & \multicolumn{1}{c}{$\hbox{Im}(\varpi) $} \\
 \hline
\multirow{2}{*}{$ 5 $} & DI  & $ -1.315457 $ & $ - 0.240587 $\\
 & CF &   $ -1.315457  $ & $ -0.240587 $  \\
\hline
\multirow{2}{*}{$ 6 $} & DI  & $ -1.590645 $ & $ -0.246052 $\\
 & CF  & $ -1.590645   $ & $ - 0.246052 $  \\
\hline
\multirow{2}{*}{$ 7 $} & DI & $ -1.865445 $ & $-0.250641  $\\
 & CF &  $ -1.865445  $ & $ - 0.250641 $ \\
\hline
\multirow{2}{*}{$ 8 $} & DI  & $ -2.139994 $ & $ - 0.254608  $\\
 & CF  & $ -2.139994   $ & $ - 0.254608 $ \\
\hline
\end{tabular*}
\begin{tabular*}{0.4\textwidth}{ c c @{\extracolsep{\fill} } c c}
\hline
 \multicolumn{4}{c}{$x_{\rm min}=0.7 $ (inside the ergoregion)} \\
\hline
 \multicolumn{1}{c}{$m$} & \multicolumn{1}{c}{Method}  & \multicolumn{1}{c}{$\hbox{Re}(\varpi) $} & \multicolumn{1}{c}{$\hbox{Im}(\varpi) $} \\
 \hline
\multirow{1}{*}{$ 5 $} & DI&  $ +1.375559  $ & $  +9.448427 \times 10^{-7} $  \\
\hline
\multirow{1}{*}{$ 6 $} & DI& $ +1.985322 $ & $   +2.932041 \times 10^{-7} $  \\
\hline
\multirow{1}{*}{$ 7 $} &DI&  $ +2.625150  $ & $  +8.056820 \times 10^{-8} $  \\
\hline
\multirow{1}{*}{$ 8 $} &DI&  $ +3.287623  $ & $  +2.081309 \times 10^{-8} $  \\
\hline\hline
\end{tabular*}
\end{center}
\label{tab-freq_1}
\end{table}

\begin{figure}[htpb!]
\includegraphics[width=0.5\textwidth]{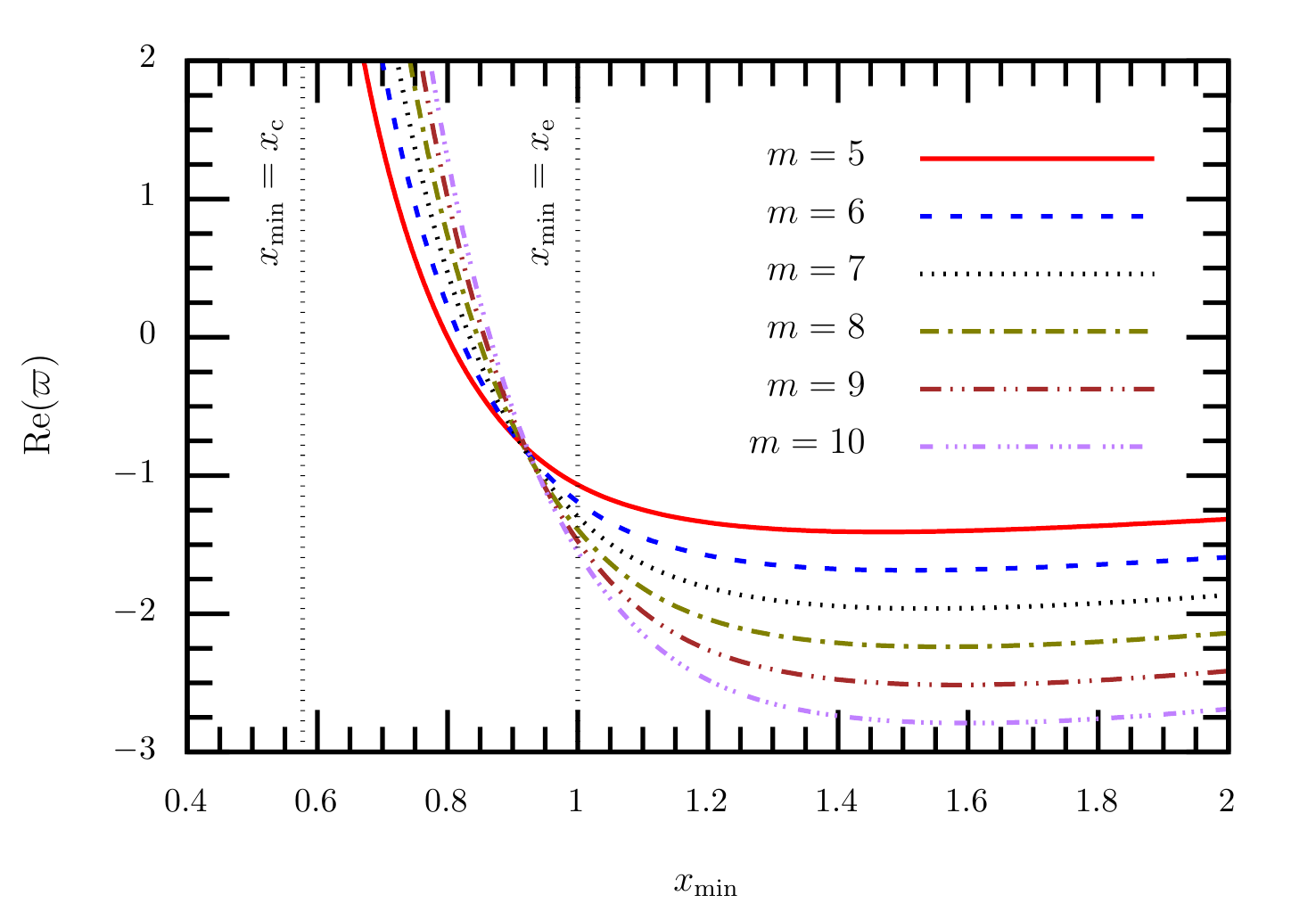}
\caption{Real part of the fundamental ($n = 0$) QNM frequencies $\varpi$ of the BEC hydrodynamic vortex, for azimuthal numbers $m=5,\,6,\,7,\,8,\,9$ and $10$, as a function of $x_{\rm min}$, obtained via DI and CF methods~\cite{Note_1}. At $x=x_{\rm min}$, we imposed a boundary condition of the Neumann type, given by Eq.~\eqref{BCII}, and at spatial infinity we considered the asymptotic behavior given by Eq.~\eqref{Boun_2}.}
\label{fig-Freq_BCII}
\end{figure}

\begin{figure*}[htpb!]
\includegraphics[width=0.5\textwidth]{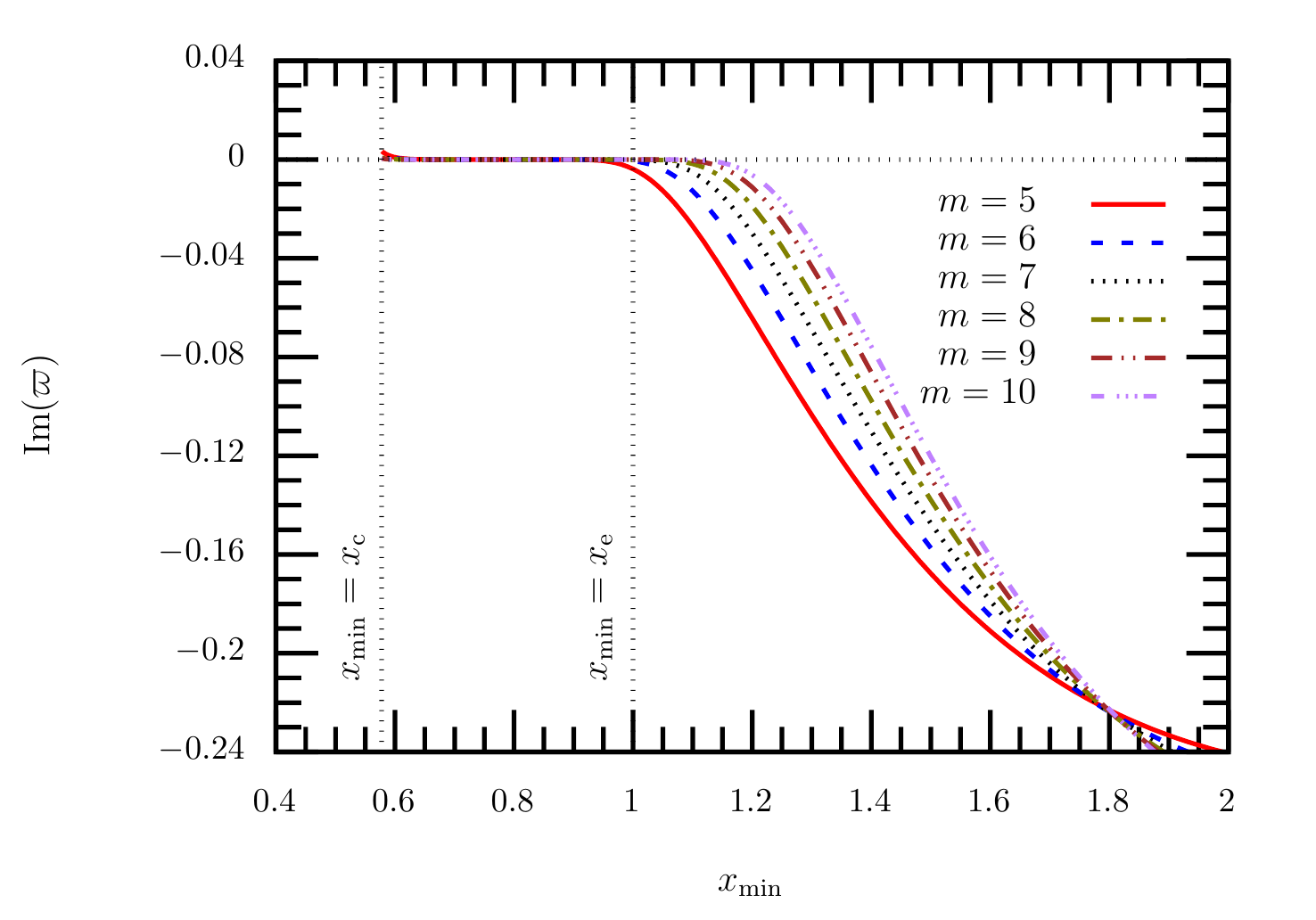}\includegraphics[width=0.5\textwidth]{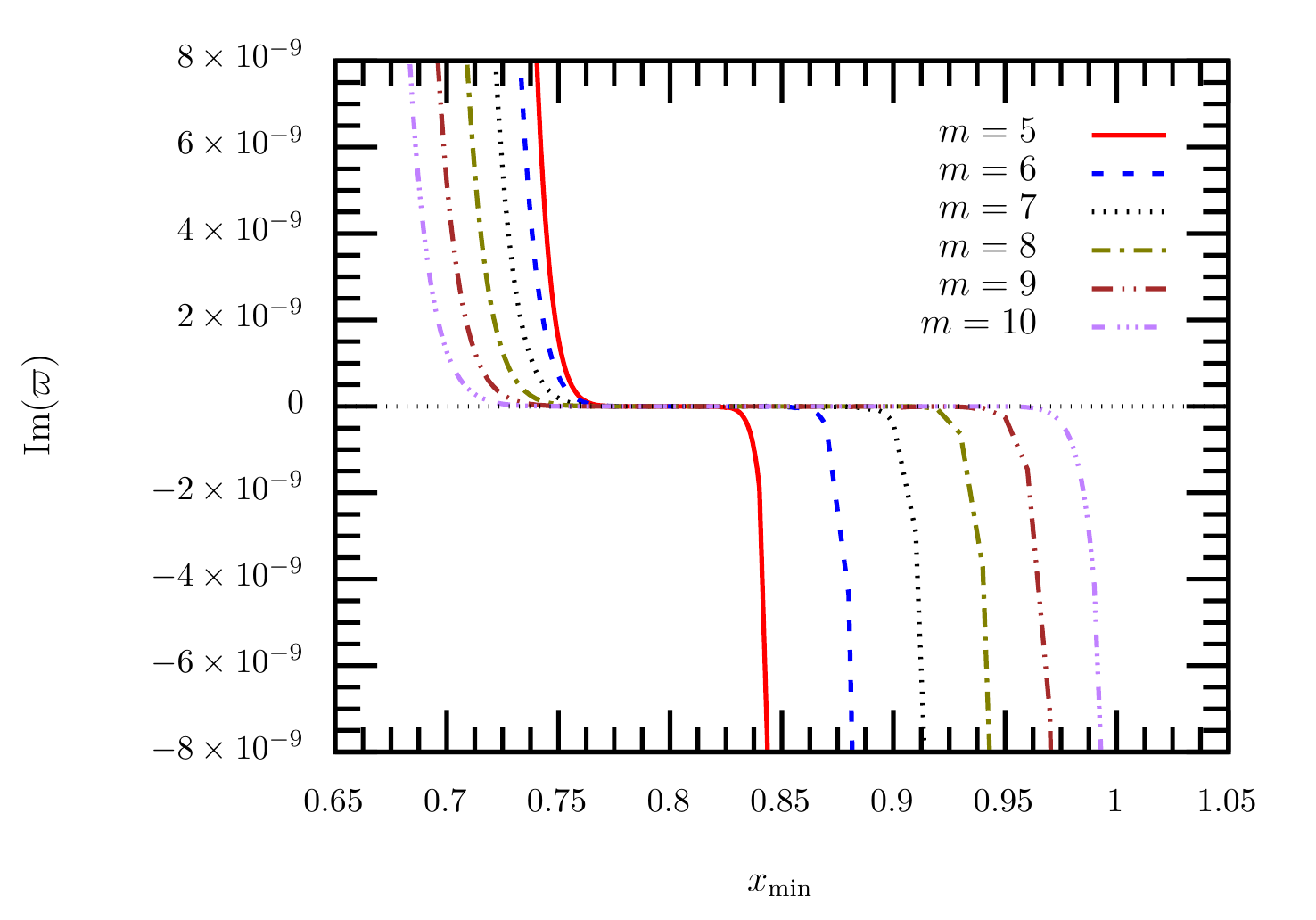}
\caption{Imaginary part of the fundamental ($n = 0$) QNM frequencies $\varpi$ of the BEC hydrodynamic vortex (left frame, with a zoom in the right frame), for azimuthal numbers $m=5,\,6,\,7,\,8,\,9$ and $10$, as a function of $x_{\rm min}$, obtained via DI and CF methods~\cite{Note_1}. At $x=x_{\rm min}$, we imposed a boundary condition of the Neumann type, given by Eq.~\eqref{BCII}, and at spatial infinity we considered the asymptotic behavior given by Eq.~\eqref{Boun_2}.}
\label{fig-Freq_BCII1}
\end{figure*}

In Figs.~\ref{fig-Freq_BCII} and~\ref{fig-Freq_BCII1} we plot, respectively, the real and imaginary parts of the fundamental ($n = 0$) QNM frequencies $\varpi$, for azimuthal numbers $m=5,\,6,\,7,\,8,\,9$ and $10$, obtained via DI and CF methods~\cite{Note_1}. We observe a decrease of the magnitude of the real and imaginary parts of the QNM frequencies for stable modes and an increase for the unstable modes. From Fig.~\ref{fig-Freq_BCII}, we may observe that there exists a certain point (around $x_{\rm min} \approx 0.9$), in which the real part of the QNM frequencies is the same for all azimuthal numbers $m$. 
We should point out that this pattern of coincidence does not happen if we, instead of the boundary condition of Neumann type, choose boundary condition of Dirichlet type.
We also note that, as we can clearly see in the right plots of Fig.~\ref{fig-Freq_BCII1}\,, the threshold between stability and instability is smaller for BEC hydrodynamic vortex than for the polytropic hydrodynamic vortex (with setup described in Ref.~\cite{Oliveira:2015vqa}), i.e., the transition from stability to instability occurs more rapidly for the BEC hydrodynamic vortex than for the polytropic hydrodynamic vortex with a compatible experimental setup in a perfect fluid (note that the transition of stability to instability is more sudden for azimuthal number $m=5$ than for $m\geq6$). 

\section{Conclusion}
\label{sec-Conclusion}
We investigated the ergoregion instability of a purely circulating system of ideal fluid (representing a quantum system): the BEC hydrodynamic vortex. We have shown that, by imposing boundary conditions inside the ergoregion of this purely circulating system, instabilities appear, which are associated with the existence of an ergoregion (supersonic flow regime) and absence of an event horizon~\cite{Oliveira:2014oja, Oliveira:2015vqa, Oliveira:2016adj}. From the QNM analysis, we concluded that the imaginary part of the QNM frequencies of this system is positive when the boundary conditions are imposed sufficiently inside the ergoregion. Furthermore, we have shown that, as the position of the boundary condition is placed more inside of the ergorigion, the system is more unstable, with the transition from stability to instability being more sudden for the BEC hydrodynamic vortex than for the polytropic hydrodynamic vortex with a compatible experimental setup in a perfect fluid~\cite{Oliveira:2015vqa}. We have thus verified for a quantum system a relevant property associated with effective spacetime with ergoregion and without an event horizon, namely the ergoregion instability.

\section*{Acknowledgments}
We thank Vitor Cardoso for the scientific collaboration that motivated the present study. 
The authors would like to thank Conselho Nacional de Desenvolvimento 
Cient\'\i fico e Tecnol\'ogico (CNPq), and Coordena\c{c}\~ao 
de Aperfei\c{c}oamento de Pessoal
de N\'\i vel Superior (CAPES), in Brazil, for partial financial support.
Financial support was also provided by the Spanish MINECO through the projects FIS2014-54800-C2-2-P and  FIS2017-86497-C2-2-P (with FEDER
contribution).
%



\begin{thebibliography}{99}
\bibitem{Note_1}
We used only the DI method to compute QNM frequencies $\varpi$ 
for $x_{\rm min} < 1$, since that CF method does not converge sufficiently well in this range.
\end{thebibliography}
\end{document}